\documentstyle[12pt,epsf]{article}

\newcommand{\ds}{\displaystyle}
\newcommand{\ra}{\rightarrow}
\newcommand{\be}{\begin{equation}}
\newcommand{\ee}{\end{equation}}
\newcommand{\bea}{\begin{eqnarray}}
\newcommand{\eea}{\end{eqnarray}}
\newcommand{\ci}{\cite}
\newcommand{\bi}{\bibitem}
\newcommand{\nono}{\nonumber \\}

\newcommand{\dd}{\partial}

\newcommand{\s}{\sigma}

\newcommand{\rr}{\vec{\bf{r}}}
\newcommand{\kk}{\vec{\bf{k}}}
\newcommand{\qq}{\vec{\bf{q}}}
\newcommand{\cx}{\mbox{\bf{\^r}}}
\newcommand{\ck}{\mbox{\bf{\^k}}}

\def\dal{\,\lower0.3ex\vbox{\hrule\hbox{\vrule\kern2pt\vbox{\kern4pt\kern4pt}
\kern2pt\vrule}\hrule}\,}

\def\s{\sigma}

\begin{document}

\title{\sl Diffraction of wave packets in space and time in three dimensions}
\vspace{1 true cm}
\author{G. K\"albermann$^*$
\\Soil and Water dept., Faculty of
Agriculture, Rehovot 76100, Israel}
\maketitle

\begin{abstract}

A simple formula for the scattering of wave packets from
a square well at long times is derived.
The expression shows that
the phenomenon of wave packet diffraction in space and time 
exists in three dimensions also.
An experiment for the verification of the phenomenon is described and simulated 
numerically.
\end{abstract}


$^*${\sl e-mail address: hope@vms.huji.ac.il}

\newpage
\section{\label{intro}\sl Introduction}

In a previous work\ci{k1}, it was shown analytically that
there exist a broad class of phenomena occurring in nonrelativistic and
relativistic quantum wave packet potential scattering
that behave as time dependent {\sl persistent} diffraction patterns.
The patterns are produced by a time independent well or barrier.
Differing form the {\sl diffraction in time}\ci{mosh} and
{\sl diffraction in space and time}\ci{zeilinger} phenomena, 
that apply to the time dependent opening of slits with
 plane monochromatic waves, the pattern does not decay exponentially with time.

Diffraction in time\ci{mosh}, and diffraction in space and time\ci{zeilinger}
are very valuable tools to investigate 
time dependent predictions of quantum mechanics.
Recent measurements of atomic wave diffraction\ci{prl},
support the theoretical prediction of the {\sl diffraction in time}
process.
The phenomenon of wave packet diffraction in space and time is still untested.
We will propose below an experiment for its verification
together with a numerical simulation of the expected results.

In ref.\ci{k1} a closed analytical formula for the
wave function of a packet scattering in one dimension off 
a time independent potential was derived.
This work was preceded by numerical 
investigations of the one and two-dimensional cases.\ci{k2,k3}
It was found analytically and numerically that, 
a diffractive structure that travels in time,
 exists for all packets, but, it survives 
only for packets that are initially narrower than a value related to the
potential effective extension. For wider packets, the peak structure
merges into a single peak.

In the present work we develop an analytical formula for the
three dimensional case and show that the same behavior is present.
In section 3 we also treat the one dimensional Dirac case
in an analytical manner.
Section 4 proposes an experiment and simulates its expected results.

\section{\label{three}\sl Wave packet diffraction in three dimensions}

Consider a gaussian wave packet, 

\bea\label{packet}
\phi(\rr,t)~&=&~e^{y_0}\nono
y_0&=&{i~\qq_0\cdot~(\rr-\rr_0)-\frac{(\rr-\rr_0)^2}{4~\s^2}}
\eea

with $\qq_0$, the initial average momentum of the packet, $\rr_0$,
 the initial location and $\s$ the packet width, impinging on a spherically
symmetric square well

\bea\label{sqwell}
V(x)= -V_0~\Theta(w-r)
\eea

where, {\sl w} is the well  width, and $V_0$, the depth, and $\Theta$, the
step function.
The results are not dependent on the choice of wave packet\ci{k1}, 
however, the Gaussian profile facilitates the integrations.
The total wave function with outgoing boundary condition reads\ci{gol}

\bea\label{psi}
\psi^+(\rr,t)~&=&\psi_{in}(\rr,t)+\psi_{scatt}(\rr,t)\nono
&=&\int{~d^3k~f(\kk,\qq_0)~(\psi_{in}(\kk,\rr,t)+\psi_{scatt}(\kk,\rr,t))}
\eea

where

\bea\label{psip}
\psi_{in}(\kk,\rr,t)&=&e^{i~\kk\cdot~\rr}\nono
\psi_{scatt}(\kk,\rr,t)&=&\frac{e^{i~k~r}}{k~r}\sum_L{(2~L+1)~e^{i\delta_L}
~sin(\delta_L)~P_L(\cx\cdot\ck)}\nono
f(\kk,\qq_0)&=&e^{-(\kk-\qq_0)^2~{~\s^2}-i~\kk~\rr_0-i~\frac{k^2}{2~m}~t}
\eea

The incoming wave may be found explicitly to be

\bea\label{psin}
\psi_{in}&=&\bigg(\frac{\pi}{it/(2m)+\s^2}\bigg)^{\frac{3}{2}}~e^{y_1}\nono
y_1&=&\frac{(\rr-\rr_0)^2-4~i~(\rr-\rr_0)\cdot~\qq_0\s^2-4\s^4~q_0^2}
{i~t/(2m)+\s^2}
\eea

Performing the angular integral over $\ck$, the scattered wave becomes

\bea\label{scatt}
\psi_{scatt}(\rr,t)&=&\int{k^2~dk~\Phi(k,t)}\nono
\Phi(k,t)&=&\frac{e^{i~k~r}}{k~r}\sum_L{(2~L+1)~e^{i\delta_L}
~sin(\delta_L)~I_L}~e^{-\s^2~(k^2+q_0^2)-i~t\frac{k^2}{2m}}\nono
\eea

The expression for $I_L$ is quite involved. However, for $L=0$ it
simplifies considerably.
Fortunately, this is the only contribution we need at long times
(except for very specific cases for which the phase shift of higher partial 
waves dominates, see below).
Fot $t\ra\infty$ the integral above oscillates wildly except for
values of {\sl k} that give an extremal phase\ci{bohm}.
It is easy to show that in this case, only extremely small values of k
enter the integral. However, for very small values of {\sl k}, the phase
shift reads\ci{gol}

\bea\label{phase}
tan({\delta}_l)&=&-\frac{(k~w)^{2~l+1}}{(2~l-1)!!~(2~l+1)!!
}~\frac{z_l-l}{z_l+l+1}\nono
z_l&=&\frac{x~j'_l(x)}{j_l(x)}
\eea

where $\ds x=k'~w=\sqrt{k^2+2~m~|V_0|}~w$, 
and $j_l$, denotes the spherical Bessel function of order {\sl l}.

It is then clear, that, except for very pathological cases only
the lowest partial wave matters.
Higher partial waves are suppressed by factors of the form 
$\ds\bigg(\frac{w}{\sqrt{it/(2m)+\s^2}}\bigg)^l$. 
For very long times we can then approximate the scattered wave by taking only
the $L=0$ term in eq.(\ref{scatt}).
For this case $I_L$ becomes

\bea\label{il}
I_0&=&4\pi~j_0(k~d)\nono
d&=&|\rr_0+2~i~\s^2\qq_0|
\eea

where the absolute value pertains to the vectorial character
of the variable inside the vertical bars, and not to the complex number.
Therefore, $j_0$ is really a complicated mixture of Bessel and
Hankel functions. 

Inserting eqs.(\ref{il},\ref{phase}) in eq.(\ref{scatt}) and recalling the
rules for the integration of the error function for complex variables, 
we find the transparent result

\bea\label{scatt1}
\psi_{scatt}&=&-\bigg(\frac{\pi}{it/(2m)+\s^2}\bigg)^{\frac{3}{2}}~
\frac{r+d}{r}~e^{y_2}\nono
y_2&=&{-\frac{(r+d)^2}{4~(it/(2m)+\s^2)}+\lambda}
\eea

where $\ds ln(\lambda)=\frac{z_0}{d~(z_0+1)}$,
with $z_l$ defined in eq.(\ref{phase}).

The incoming and scattered wave interfere in the total wave.
Noting that the scattered wave depends on the absolute value of $\rr$,
while the incoming wave depends on $\rr$ itself, it is clear
that there will be a totally different behavior at forward and backward angles.
The backwards scattered full wave for long distances and times becomes

\bea\label{back}
\psi&=&2~i\bigg(\frac{\pi}{it/(2m)+\s^2}\bigg)^{\frac{3}{2}}
~sin(\frac{m~r}{t}~(r_0+2~i~q_0\s^2)+\lambda/2)~e^x\nono
x&=&{-q_0^2\s^2-\frac{r^2~m^2}{t^2}+i\frac{t}{2~m}~r^2+\lambda/2}
\eea

This is a diffraction pattern that travels in time as found for the
one dimensional case\ci{k1}. Again the pattern gets blurred and 
forms a single peak when the imaginary part of the argument of the
{\sl sin} function is large. As explained in ref.\ci{k1}, this
amounts to the condition $\s>>\sqrt{\frac{w}{q_0}}$.
For narrow enough packets the pattern persists to infinity.
In the forward direction there is no interference as may be easily
found by examining eq.(\ref{scatt1}).
Instead of a {\sl sin} function, we find there a pure exponential.
Only tiny interferences survive due to small terms.
In some sense the process should be more properly called, interference
in space and time, because it is due to the superposition of incoming
spread packet and the scattered wave. 

Numerical calculations of the two dimensional case\ci{k3} indeed
support these findings.
Moreover, the results are independent of the impact parameter
provided the initial packet is located at a much longer distance along the
direction of $\qq_0$ as compared to the transverse direction, 
as found numerically in ref.\ci{k2}.
Also the pattern does not depend strongly on the initial
momentum.
The formula of eq.(\ref{back}) may be used to find the angular
distribution, the scattering and total cross sections, etc.
Such expressions do not seem to exist in the literature for packets, even
for the simplest square well case dealt with here.

As mentioned above the above treatment took advantage of the
properties of Gaussian packets and the simplicity of the square well.
However, they are general. One has only to replace the Gaussian form factor
by the appropriate one in case and the phase shift for
the square well by the one corresponding to the specific choice of potential.
The case of long range potentials deserves however some care.
In refs.\ci{k1,k2,k3} the connection to the excitation of a half-bound
state, inside the well, Levinson's theorem and the
phenomena of diffraction in time and diffraction in space and time
 were discussed. The same applies without modification to the
present case.

It is also evident, that the sign of the potential is irrelevant.
A barrier may do too.\ci{k1} This is in line with optical diffraction
of slits and around objects.

\section{\label{dirac}\sl Wave packet scattering for the Dirac equation}

Consider the one-dimensional Dirac equation with a scalar potential
 {\sl S(x)} as given by eq.(\ref{sqwell})

\bea\label{Dirac}
\bigg[i\gamma^{\mu}~{{\dd_{\mu}}}+\big(m+S(x)\big)\bigg]\Psi=0
\eea

For the initial packet we use a minimal uncertainty relativistically
invariant wave packet

$\Psi_0 =\left( \begin{array}{c}U \\ V\end{array} \right)$

\bea\label{dirpacket0}
U&=&\int{dk~e^{i~k(x-x_0)-2~\s^2~(E~E_0-k~q_0-m^2)}}\nono
V&=&\int{dk~\frac{i~k}{E+~m}~e^{i~k(x-x_0)-2~\s^2~(E~E_0-k~q_0-m^2)}}
\eea

with $E=\sqrt{k^2+m^2},~E_0=\sqrt{q_0^2+m^2}$, and we have taken 
$\bf \ds \gamma_0=\bf \sigma_z~,\gamma_1=\sigma_z~\sigma_x$ for the Dirac 
matrices, where $\sigma$ denotes the corresponding Pauli matrix.

The stationary wave function inside and outside the well for fixed momentum
is given by $\phi(k,x,t)
 =\left( \begin{array}{c}\phi_1 \\ \phi_2\end{array} \right)$

\bea\label{psidir}
\phi_1(x<-w,k,t)&=&~(e^{i~k~x}+~B~e^{-i~k~x})\nono
\phi_2(x<-w,k,t)&=&\frac{i~k}{E+m}~(e^{i~k~x}-~B~e^{-i~k~x})\nono
\phi_1(-w<x<w,k,t)&=&~(C~e^{i~k'~x}+~D~e^{-i~k'~x})\nono
\phi_2(-w<x<w,k,t)&=&~\frac{i~k'}{E+m^*}~(C~e^{i~k'~x}-~D~e^{-i~k'~x})\nono
\phi_1(w<x,k,t)&=&~F~e^{i~k~x}\nono
\phi_2(w<x,k,t)&=&~\frac{i~k}{E+m}~F~e^{i~k~x}
\eea
with

\bea\label{dirpar}
B&=&\frac{(1-g^2)(e_3^2-e_4^2)}{\Delta}\nono
C&=&\frac{-2~e_1~e_4~(1+g)}{\Delta}\nono
D&=&\frac{2~e_1~e_3~(1-g)}{\Delta}\nono
F&=&\frac{-4~g}{\Delta}
\eea

where $\ds e_1=e^{i~k~w},~e_2=e^{-i~k~w},~e_3=e^{i~k'~w},~e_4=e^{-i~k'~w}$,
 $\ds~g=\frac{k'~(E+m)}{k~(E+m^*)}$, $\ds \Delta=e_3^2~(1-g)^2-e_4^2~
(1+g)^2$, $\ds k'=\sqrt{E^2-(m^*)^2},~m^*=m-V_0$.\footnote
{For a vector potential the $^*$ is in the energy instead of the mass}

The solution to the scattering of the packet becomes
$\Psi(x,t) =\left( \begin{array}{c}U(x,t) \\ V(x,t)\end{array} \right)$

\bea\label{dirpacket}
U(x,t)&=&\int{dk~\phi_1(k,x,t)
~e^{-i~k~x_0-2~\s^2~(E~E_0-k~q_0-m^2)-i~E~t}}\nono
V(x,t)&=&\int{dk~\phi_2(k,x,t)
~e^{-i~k~x_0-2~\s^2~(E~E_0-k~q_0-m^2)-i~E~t}}
\eea

As in section \ref{three} we consider the long time behavior of the 
wave. Here too, the reflected wave will receive contributions mainly
from low values of k.
Inspection of the reflection parameter {\sl B} in eq.(\ref{dirpar}), we
find that for $k\ra~0$, $B\ra~-1$.
The same happened in the nonrelativistic treatment of \ci{k1}.
For low initial momenta, $q_0<<m, or~v<<1$, the exponential of 
eq.(\ref{dirpacket}) becomes
$\ds \approx~e^{-~ik~x_0-i~m~t-i~\frac{k^2}{2~m}t-\s^2~(q_0-k)^2}$.
So except for the factor $\ds e^{i~m~t}$, that is independent of {\sl k},
we are back at the nonrelativistic case for the upper component, and at the
same time the lower component vanishes.
Hence, the Dirac case in one dimension is exactly analogous to the 
nonrelativistic case for low average momenta of the initial wave packet.

In the ultrarelativistic limit $\ds m\ra~0$, the upper and lower components are
identical up to a factor of {\sl i}. 
The incoming and reflected waves do not spread. Hence, there
is no interference between both. 
If we first find the wave in the backward direction as in ref.\ci{k1}, and then
take the limit of a small mass, we find that the pattern disappears, as
the arguments of the {\sl sin} function tend to zero.
For $t\ra\infty$, the backwards scattered wave
 becomes a receding wave packet located
around $x=-x_0-t-w$, without any diffraction. The incoming packet is no
longer present in the region behind the well.

In summary, the diffraction in space and time
with wave packets exists in the relativistic regime, but gradually
diminishes as the mass decreases, until 
is eventually washed out completely for the massless case, as was
 found for the {\sl diffraction in time} process with the
Klein-Gordon equation.\ci{mosh}

\section{\label{experiment}\sl Numerical simulation of a suggested experiment}

Present advances in bose traps permit the easy handling of atoms at very low
temperature and velocities. At such energies the atoms move as if they
were a wave packet provided the random agitation due to thermal effects
is not as crucial.
However, still the large amount of interactions
between the particles and of them with the environment 
causes a decoherence of the wave in quite a short time.

Instead of resorting to a cold bose gas, we will then focus
on a drop of liquid Helium.
Although the experimental details are beyond the expertise of the
present theoretical work we will describe a hypothetical setup
in the hope that it is not too far away from reality.

Consider a drop of liquid helium, of around $1~cm^3$ in volume
lying at a corner of Cesium coated plates perpendicular to
each other. Cesium is needed in order to prevent wetting\ci{nacher}.
The helium, plate
and environment should be  at a temperature
below the $\lambda$ point, although it would be
interesting to see the effect of normal to superfluid component
of the Helium on the wave packet behavior, as a tool to study decoherence.
Put a large area detector behind the drop behind the plate.
The number of particles measured as a function of time at a
fixed position would then be determined by the absolute value of
the wave function at that point. The particles are repelled
by the wall and the packet spreads at the same time.
Figure 1 depicts such a case. We solved the one dimensional Schr\"odinger
equation for a packet as described in \ci{k1} 
, with and a barrier of very large strength.
We assumed a total number of particles of $N=5*10^{21}$, obtained from
the tabulated density of liquid helium and a volume of 1 $cm^3$.
We used a plate (potential) of width w=1 cm and strength $V_0=4~eV\approx
10^{15}$, in units of $\hbar=1$. We took a detctor size of $dx=1~mm$.
It appears that, due to the large number of particles to be detected a
simple weighing technique might be feasible, in this way we also
avoid interaction of the packet with the detector.

\begin{figure}
\epsffile{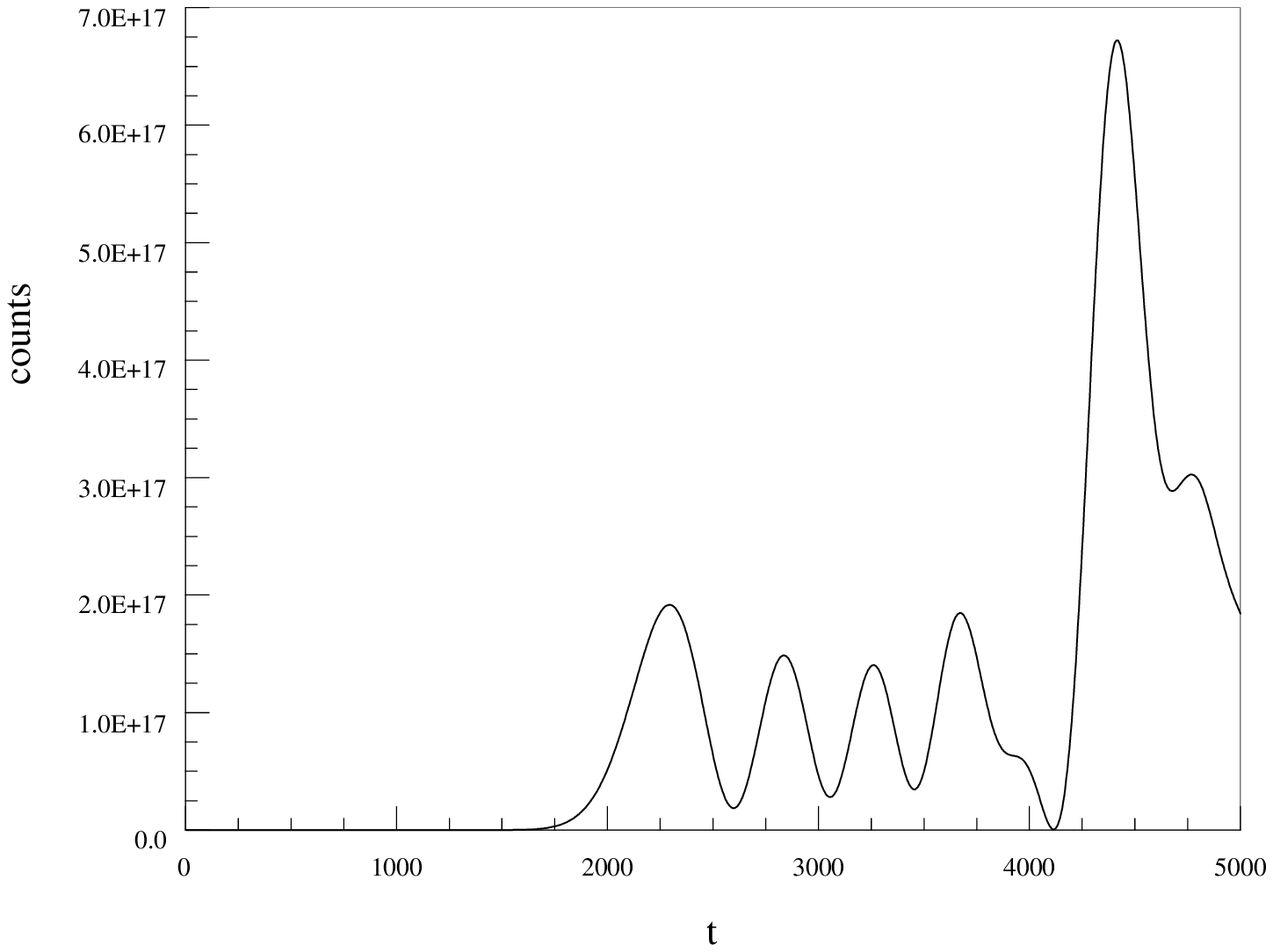}
\vsize=5 cm
\caption{\sl Number of Helium atoms counted at a distance of 5 cm
behind a plate of thickness w=1 cm as a function of time in seconds}
\label{fig1}
\end{figure}

It is perhaps too optimistic to expect the proposed experiment will work
so cleanly as in the simulation, due to all kinds of effects inside the
drop that lead to decoherence as well as production of
internal excitations like rotons, vortons, etc.
However, some remainder of the effect might still
show up in the counter.
If it does, it will be a triumph for the quantum mechanical description
of matter waves by means of wave packets of macroscopic size.

\end{document}